# Wave mechanics without gauge fixing

## A. M. Stewart

Research School of Physical Sciences and Engineering,

The Australian National University,

Canberra, ACT 0200, Australia.

**Abstract.** By using the multipolar gauge it is shown that the quantum mechanics of an electrically charged particle moving in a prescribed classical electromagnetic field (wave mechanics) may be expressed in a manner that is gauge invariant, except that the only gauge functions that are allowable in a gauge transformation are those that consist of the sum of a function that depends only on the space coordinates and a term that is the product of a constant and the time coordinate. The multipolar gauge specifies the specific set of potentials that are best suited for use in the Schroedinger equation.

---------------------------

This paper discusses the interpretation of the Schroedinger equation for a single charged particle minimally coupled to an externally prescribed classical electromagnetic field. It is shown that when the multipolar gauge is used the correct potentials to be used in the Schroedinger equation emerge automatically and do not have to be put in by hand. Not all gauge transformations may be used legitimately in wave mechanics and those that are allowed to be used are found.

The Schroedinger equation for a particle of charge *e* and mass *m* moving in an externally prescribed classical electromagnetic field described by vector and scalar potentials $A(r,t)$ and $\phi(r,t)$ is

$$H\Psi(r,t) = i\hbar (\partial/\partial t)\Psi(r,t) \qquad (1)$$

where the Hamiltonian $H = (p - eA)^2/2m + e\phi$, the canonical momentum becomes $p = -i\hbar\nabla$ and $\Psi(r,t)$ is the wavefunction.

If a gauge transformation is made of the potentials

$$A \to A_\chi = A + \nabla\chi, \qquad \phi \to \phi_\chi = \phi - \partial\chi/\partial t \qquad (2a)$$

and of the wavefunction

$$\Psi(r,t) \to \Psi_\chi(r,t) = \Psi(r,t)\exp[ie\chi(r,t)/\hbar] \qquad (2b)$$

then the Schroedinger equation (1) is invariant in form under the transformation and the charge and current densities are unchanged. The arbitrary scalar field $\chi(r,t)$ is known as the gauge function and is required to satisfy the condition $[(\partial/\partial i)(\partial/\partial j) - (\partial/\partial j)(\partial/\partial i)]\chi = 0$ everywhere, where *i* and *j* are any pair of the coordinates *x, y, z* and *t*.

This form invariance of the Schroedinger equation under a gauge transformation is sometimes taken to imply that the predictions of wave mechanics are themselves gauge invariant, but such an assumption is not warranted. To demonstrate that the predictions of wave mechanics do not depend on gauge it is necessary to show explicitly that all measurable properties and transition probabilities do not depend on gauge either.

If an operator *V* transforms as

$$V_\chi = \exp(ie\chi/\hbar) V_0 \exp(-ie\chi/\hbar) \qquad (3)$$

under a gauge transformation, where the subscript denotes the gauge that is used, 0 denoting a reference gauge, then it is clear that its matrix elements, which determine measurable quantities, will be independent of gauge because the exponents in the operator and wavefunction (2) containing the gauge functions cancel. It is generally accepted that any measurable prediction of the theory must be independent of the gauge function that is used and that operators corresponding to physically measurable quantities must have the form of equation (3). However, as noted as early as 1937 [1] (a translation by D. ter Haar from the Dutch of the 1937 book of Kramers), the Hamiltonian of the Schroedinger equation, which is expected to represent the physically observable energy, does not obey equation (3), transforming instead as

$$H_\chi = \exp(ie\chi/\hbar) H_0 \exp(-ie\chi/\hbar) - e\, \partial\chi/\partial t \qquad (4)$$

The concept of energy requires that for a conservative system (fields are independent of time) the energy be constant in time. Taking the matrix elements of equation (4) between the states $|n_\chi\rangle = \psi_n(\mathbf{r})\exp[ie\chi(\mathbf{r},t)/\hbar]$, where the $\psi_n$ are *any* orthonormal set of basis states we get

$$\langle m_\chi | H_\chi | n_\chi \rangle = \langle m_0 | H_0 | n_0 \rangle - e \int \psi_m^*(\mathbf{r})[\partial\chi(\mathbf{r},t)/\partial t]\psi_n(\mathbf{r})\,d\mathbf{r}. \quad (5)$$

If the system is conservative, then by definition, in zero gauge the matrix element $\langle m_0|H_0|n_0\rangle$ is independent of time. In order for the gauge-arbitrary matrix element of equation (5) to be independent of time as well we require that $\partial^2 \chi(\mathbf{r},t)/\partial t^2 = 0$. The appropriate solution to this equation is

$$\chi(\mathbf{r},t) = f(\mathbf{r}) + gt \quad (6)$$

where $f(\mathbf{r})$ is an arbitrary scalar function of $\mathbf{r}$ and $g$ is an arbitrary constant independent of both $\mathbf{r}$ and $t$. The quantity $g$ must be independent of $\mathbf{r}$ because if it did depend on $\mathbf{r}$ then the off-diagonal matrix elements of (5) could be changed by a gauge transformation and probability amplitudes would become gauge dependent. Equation (6) exhibits the form of the only gauge functions that are allowed to be used in wave mechanics. The consequence is that, under an acceptable gauge transformation, all diagonal matrix elements of the Hamiltonian are shifted by a constant energy (- *eg*) so that energy *differences* are unchanged and off-diagonal matrix elements are left unchanged also.

  As has been argued before [2], it follows from these considerations that wave mechanics, or semi-classical electrodynamics, is not a fully gauge invariant theory. The reason for this is that, unlike classical and quantum electrodynamics, which *are* fully gauge invariant theories wave mechanics is not a properly formulated theory of the interaction of particles and fields. It takes account of the energy of the particle and of the interaction between the particle and the field but does not take account of the energy of the field. Consequently energy is not conserved in the exchange of energy between particles and fields.

One example of a gauge transformation that, on these grounds, is not acceptable in wave mechanics is the transformation to the temporal gauge. As noted by Kobe [3] by making a transformation with gauge function $\chi(\boldsymbol{r},t) = \int_T^t \phi(\boldsymbol{r},t')\mathrm{d}t'$ the transformed scalar potential becomes zero and the Hamiltonian is entirely kinetic. However, according to equation (6) this transformation is not allowed in wave mechanics, although it is in a fully gauge invariant theory such as field theory.

The limitation to the gauge functions given by equation (6) seems, in practice, to not have given rise to any problems in the quantum chemistry literature. Gauge functions that are not of that form give obvious trouble immediately in wave mechanics and, to the author's knowledge, have never been used in that subject, although they are common in quantum field theory. Accordingly, previous calculations of such properties as magnetic susceptibilities, chemical shifts etc, are unlikely to have been in error from this cause.

If a theory is gauge invariant, it does not matter what gauge is used to express the potentials, but if the theory is not gauge invariant then it does matter. It was argued before [2] that, in order for the scalar potential to represent the physical potential energy, the appropriate gauge to use was the multipolar gauge [2, 4-9], known also by a multitude of other names [10]. Further, to actually compute a matrix element of the potentials in equation (1), the gauge has to be "fixed", i.e. the gauge function or its gradients have to be set to zero. It then must be shown that all results turn out to be independent of the actual value of the gauge function, whatever it is taken to be.

In this note we also show how the use of the multipolar gauge enables suitable potentials to be obtained and a gauge invariant treatment of wave mechanics to be made. The restriction is made that only the gauge functions of equation (6) are allowed. We consider spatially localized physical systems common in atomic and molecular physics.

A *gauge transformation* is made by replacing the unspecified gauge function $\chi(\mathbf{r},t)$ in (2) by the sum of a *specified* function and another *unspecified* function. The algebra of arbitrary functions implies that the sum of two arbitrary functions is an arbitrary function and the sum of an arbitrary function and a function that is explicitly expressed in $x,y,z$ and $t$ is also an arbitrary function. The multipolar gauge is obtained in this way by means of a gauge transformation [5, 6] and expresses the potentials $\mathbf{A}$ and $\phi$ at position $\mathbf{r}$ and time $t$ in terms of the gauge invariant fields $\mathbf{E}$ and $\mathbf{B}$ as

$$\mathbf{A}(\mathbf{r}, t) = -(\mathbf{r} - \mathbf{R}) \times \int_0^1 \mathbf{B}(\mathbf{q}, t) u \, du + \nabla \chi(\mathbf{r}, t) \qquad (7a)$$

and

$$\phi(\mathbf{r}, t) = -(\mathbf{r} - \mathbf{R}) \cdot \int_0^1 \mathbf{E}(\mathbf{q}, t) \, du - \partial \chi(\mathbf{r}, t) / \partial t \qquad (7b)$$

The 3-vector $\mathbf{q}$ is given by $\mathbf{q} = u\mathbf{r} + (1 - u)\mathbf{R}$, so the integrals in equations (7) are taken along a straight line in $(x,y,z)$ space from an arbitrary reference point $\mathbf{R}$, say the nucleus of an atom, to the field point $\mathbf{r}$. This gauge, like the Coulomb gauge, is instantaneous. The presence of the arbitrary function, again called $\chi$ for simplicity, indicates that these potentials still retain their gauge freedom. The gauge only becomes *fixed* when the gradients of the arbitrary function are set equal to zero. The coupling of the charge to the electromagnetic field may therefore be expressed either as a local coupling to the potentials as in the Hamiltonian of (1) or as a non-local coupling to the fields as in (7), but in either case full gauge freedom is maintained [11, 12].

Our approach is to express the potentials in terms of the fields $E$ and $B$ that are physically real, in the sense that they are capable of being measured. We take these fields to be composed of time-independent parts and time-dependent parts, $E(r,t) = E^0(r) + E^1(r,t)$ and $B(r,t) = B^0(r) + B^1(r,t)$. These give rise, through equations (7), to the corresponding potentials $A(r,t) = A^0(r) + A^1(r,t) + \nabla\chi$ and $\phi(r,t) = \phi^0(r) + \phi^1(r,t) - \partial\chi/\partial t$, noting that the sum of two arbitrary scalars is another arbitrary scalar. The time-independent parts $A^0(r)$ and $\phi^0(r)$ determine the wavefunctions and energies of the stationary states of the system [2], the time-dependent parts $A^1(r,t)$ and $\phi^1(r,t)$ give rise to quantum transitions between these stationary states. When we say that wave mechanics is done without gauge fixing we mean that although the potentials $A^0, A^1, \phi^0$ and $\phi^1$ are fixed in the specific forms prescribed by equation (7), the potentials still retain their full gauge arbitrariness from the presence of the derivatives of $\chi$.

First we consider the time-independent potentials given by (7). The scalar potential is

$$\phi^0(r) = - \int_R^r E^0(q).dq \qquad (8)$$

For time-independent fields this integral does not depend on the path of integration and we recognize it as the work done to move the charged particle from $R$ to $r$, or from the nucleus to $r$, in other words the physical potential energy of the particle. To attain the conventional result we add to (6) a constant term that does not depend on $r$, $-\int_\infty^R E^0(q).dq$ to get $\phi^0(r) = -\int_\infty^r E^0(q).dq$, the potential referred to spatial infinity which, for an atomic system, is conventionally defined to be the atomic potential energy $\phi^a(r)$ of an electron at $r$ so that $\phi^0(r) = \phi^a(r)$. This is an example of making a gauge transformation with a gauge function consisting of a constant multiplied by the time, namely $\chi = t\int_\infty^R E^0(q).dq$. Next we consider the vector potential. If the magnetic field $B^0$ is uniform over the size of the system considered, then the integral over $u$ in (7) may be done trivially to give 1/2 and we get $A^0(r) = B^0 \times (r - R)/2$. Arbitrary non-uniform fields are dealt with by using the full form of the vector potential in equation (7) [2]. In the absence of electric currents the explicit parts of the potentials (7) obey the Lorenz gauge condition $\nabla.A + c^{-2} \partial\phi/\partial t = 0$; when the electric field is stationary as well, the Coulomb gauge condition $\nabla.A(r) = 0$ is satisfied.

Now we consider the time-dependent parts of the potentials. By expanding the *fields* about the point $R$ [6, 7] we get

$$\phi^1(r,t) = -(r - R).E^1(R,t) - \sum_{i,j}(x^i - X^i)(x^j - X^j)[\partial E^{1i}(R,t)/\partial x^j]/2 + ..., \quad (9)$$

where the *i* and *j* are the Cartesian components of the vectors, the first term having the form of an electric dipole interaction, the second that of an electric quadrupole, where the derivatives are evaluated at *r* = *R*, and

$$A^1(r,t) = -(r - R) \times B^1(R,t)/2 + \ldots \qquad . \qquad (10)$$

There are higher multipole terms that depend on the gradients of the fields in the expansions for both potentials [6] but we ignore them for systems of limited spatial extent. We have therefore found explicit and gauge invariant expressions for the potentials $A^0, A^1, \phi^0$ and $\phi^1$. Gauge-equivalent sets of potentials may be constructed by making a gauge transformation with a gauge function satisfying equation (6).

The Schroedinger equation (1) in arbitrary gauge may now be written with the Hamiltonian partitioned into two parts as [2, 13, 14]

$$[H^0_\chi + V_\chi - i\hbar(\partial/\partial t)]\Psi_\chi(r,t) = 0 \qquad (11)$$

where

$$H^0_\chi = [p - e(A^0 + \nabla\chi)]^2/2m + e(\phi^0 - \partial\chi/\partial t) \qquad (12)$$

and

$$V_\chi = e\phi^1 - A^1 \cdot [p - e(A^0 + \nabla\chi)]e/m + e^2(A^1)^2/2m + ie\hbar(\nabla \cdot A^1)/2m \qquad (13)$$

where the time-independent (with superscript 0) and time-dependent (with superscript 1) parts of the potential have been separated and the gauge function has been displayed explicitly. The unperturbed part of the Hamiltonian $H^0_\chi$ is not gauge invariant, as discussed previously. The perturbed part $V_\chi$ is gauge invariant as it satisfies equation (3).

The gauge-explicit stationary states are obtained from $[H^0_\chi - i\hbar(\partial/\partial t)]\Psi_{\chi,n}(\boldsymbol{r},t) = 0$ to be [2, 13, 14]

$$\Psi_{\chi,n}(\boldsymbol{r},t) = \psi_n(\boldsymbol{r}) \exp\{i[e\chi(\boldsymbol{r},t) - E_n t/\hbar]\} \qquad (14)$$

where it is assumed that solutions to the eigenvalue equation

$$[(\boldsymbol{p} - e\boldsymbol{A}^O)^2/2m + e\phi^O]\psi_n(\boldsymbol{r}) = E_n \psi_n(\boldsymbol{r}) \qquad (15)$$

can be found since $\boldsymbol{A}^O$ and $\phi^O$ are time-independent and well defined.

With time-dependent fields, the wavefunction becomes a time-dependent linear combination of the basis wavefunctions, the expansion coefficients being the probability amplitudes $a_{\chi,n}(t)$

$$\Psi_\chi(\boldsymbol{r},t) = \sum_n a_{\chi,n}(t)\Psi_{\chi,n}(\boldsymbol{r},t) \qquad . \qquad (16)$$

There appears to be no *prima facie* reason why the probability amplitudes should not depend on the gauge function $\chi$, although their squared moduli, being the physical probabilities, must not. By substituting equation (16) into (11) and noting that $V_\chi$ satisfies equation (3), the result is obtained that [2, 13, 14]

$$i\hbar \frac{d a_{\chi,m}(t)}{dt} = \sum_n a_{\chi,n}(t) V_{m,n}(t) \exp[i(E_m - E_n)t/\hbar] , \qquad (17)$$

where

$$V_{m,n}(t) = \int d\mathbf{r}\, \psi_m^*(\mathbf{r}) V_0(\mathbf{r}, t) \psi_n(\mathbf{r}) \qquad . \qquad (18)$$

It is seen that $V_{m,n}(t)$ comes out to be independent of gauge. Consequently the equations of motion (17) for the $a_{\chi,m}(t)$ are precisely the same for all gauges and there is no loss of generality in taking the probability amplitudes to be independent of gauge namely $a_{\chi,m}(t) = a_{0,m}(t)$. This derivation is formally correct for any gauge function but, as noted, only gauge functions that satisfy equation (6) allow the Hamiltonian to be interpreted as the energy.

The matrix element $V_{m,n}(t)$ of the perturbation comes from equation (13) with the perturbing potentials given by equations (9) and (10), so the perturbing operator is

$$V_0(\mathbf{r},t) = -e(\mathbf{r} - \mathbf{R}).\mathbf{E}^1 - (e/2m)\mathbf{l}.\mathbf{B}^1 + e^2[(\mathbf{r} - \mathbf{R})\mathbf{x}\mathbf{B}^1].[(\mathbf{r} - \mathbf{R})\mathbf{x}\mathbf{B}^0]/4m$$

$$+ e^2[(\mathbf{r} - \mathbf{R})\mathbf{x}\mathbf{B}^1]^2/8m + \ldots \qquad . \qquad (19)$$

The first term on the right is an electric dipole term, the next is a magnetic dipole term with $\mathbf{l} = (\mathbf{r} - \mathbf{R})\mathbf{x}\mathbf{p}$; this arises from the $\mathbf{A}^1.\mathbf{p}$ term in (13). The next two terms are diamagnetic ones. For typical atomic systems the terms get progressively smaller going from left to right [7]. If terms from (9) and (10) involving the field gradients are needed they will appear both in the applied fields and, for the magnetic quadrupole and higher terms, also in the $ie\hbar(\nabla \cdot \mathbf{A}^1)/2m$ term in the Hamiltonian. It is seen that the multipolar gauge prescribes in a direct manner the most useful specific and exact expressions for the potentials of the Schroedinger equation.

The above discussion has dealt with a single charged particle in an external field. For a many particle system, such as an atom with many electrons, the Hamiltonian consists of the sums of single particle Hamiltonians as in equation (1) plus the instantaneous Coulomb interaction between the particles [15].

In summary, we have found that, by expressing the electromagnetic potentials of the Schroedinger equation in the multipolar gauge, explicit expressions for these potentials that are most appropriate for use in the equation emerge automatically. The predictions of wave mechanics are found to be independent of gauge provided that only gauge transformations are made that are transformations from the multipolar gauge that do not explicitly depend on time. The only exception to this rule is a transformation that has a gauge function equal to the time multiplied by a constant, as this preserves energy differences.